\documentclass{desyproc}

\usepackage{graphicx}
\begin{document}
\title{CMS results on soft diffraction}
\author{{\slshape Konstantin Goulianos}\\[1ex]
The Rockefeller University, 1230 York Avenue, New York, NY 10065, USA\\[1ex]
(On behalf of the CMS Collaboration)}

\contribID{smith\_joe}


\acronym{EDS'09} 

\maketitle

\begin{abstract}
 We present measurements of soft single- and double-diffractive  cross sections, as well as of forward rapidity gap cross sections at 7 TeV at the LHC, and compare the results to other measurements and to theoretical predictions implemented in various Monte Carlo simulations.    

\end{abstract}
\vspace*{-1em}
\section{Introduction}\label{sec:intro}
Diffractive interactions are characterized by the presence of at least one non-exponentially suppressed large rapidity gap ({\sc lrg}) in the final state, defined as a region in pseudorapidity devoid of particles. The origin of {\sc lrg}s is attributed to a color-singlet exchange with vacuum quantum numbers, referred to as Pomeron ($I\!\!P$) exchange. 

Soft diffractive interactions (with no hard scale) cannot be calculated within perturbative {\sc qcd} (p{\sc qcd}), and traditionally have been described by models based on Regge theory. Model predictions generally differ when extrapolated from pre-{\sc lhc} energies ($\sqrt{s} \le 1.96$~TeV) to 7~TeV at {\sc lhc}. Thus, measurements of diffractive cross sections at the {\sc lhc} provide a valuable input for understanding diffraction and improving the modeling of diffraction in current event generators. 

The first {\sc cms} measurements of single-diffractive (SD), double-diffractive (DD), and forward rapidity gap cross are presented. Results are based on 2010 data when the {\sc lhc} was running in a low pile-up scenario, most suitable for event selection based on the {\sc lrg} signature. 

\vspace*{-1em}
\section{Inclusive diffractive cross sections}
\begin{wrapfigure}{r}{0.5\textwidth}
\vspace{-1.5em}
\includegraphics[width=.5\textwidth]{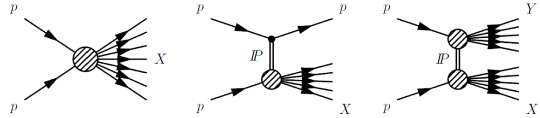}
\vspace*{-2em}
\caption{Diagrams of non-diffractive (left) and diffractive processes: single-dissociation (middle) and double-dissociation (right).}
\vspace*{-1em}
\label{fig1}
\end{wrapfigure}

\begin{figure*}[!htb]
\vspace*{-1em}
\includegraphics[width=1.0\textwidth]{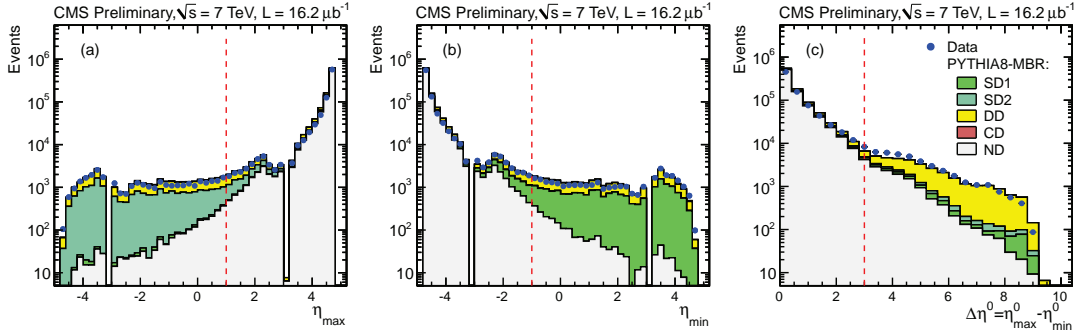}
\vspace*{-3em}  
\caption{Data distributions for (a) $\eta_{max}$, (b) $\eta_{min}$ and (c) $\Delta\eta^{0} = \eta^{0}_{max}-\eta^{0}_{min}$, compared to {\sc pythia8-mbr} predictions normalized to the luminosity of the data. The DD MC generated events are scaled downwards by 15~$\%$.}\label{fig2}
\end{figure*}

Diffractive cross sections have been measured~\cite{softdiff} based on a minimum-bias sample at $\sqrt{s}=7$ TeV. The {\sc sd} and {\sc dd} events (Fig.~\ref{fig1}) were separated using the {\sc castor} calorimeter, which covers the very forward region of the experiment, -6.6 $< \eta <$ -5.2. Minimum-bias events were selected by requiring a signal above noise level in any of the {\sc bsc} (Beam Scintillator Counter) devices ($3.2<|\eta|<4.7$) and the presence of at least two energy deposits in the central detector ($|\eta| \lesssim 4.7$). No vertex requirement was imposed. 
Diffractive events were selected by requiring the presence of a forward rapidity gap reconstructed at the edge of the central detector or central gap. The forward gap on the positive (negative) side was reconstructed in terms of the variable $\eta_{max}$ ($\eta_{min}$) defined as the highest (lowest) $\eta$ of a particle reconstructed in the detector. The central gap was reconstructed as $\Delta\eta^{0} = \eta^{0}_{max}-\eta^{0}_{min}$, with $\eta^{0}_{max}$ ($\eta^0_{min}$) defined as the closest-to-zero $\eta$ of a particle reconstructed on the positive (negative) $\eta$-side of the central detector, with an additional requirement of activity on both sides of the detector. Figure~\ref{fig2} shows distributions of $\eta_{max}$, $\eta_{min}$ and $\Delta\eta^{0}$ compared to predictions of {\sc pythia8-mbr}~\cite{mbrnote,dinoModel}. Diffractive events appear as a flattening of the exponentially falling distributions of non-diffractive ({\sc nd} events  with rapidity gaps due to random multiplicity fluctuations, and dominate the regions of low $\eta_{max}$, high $\eta_{min}$, or high $\Delta\eta^{0}$. 
\begin{figure*}[!htb]
\vspace*{-1 em}
\begin{center}
\includegraphics[width=1\textwidth]{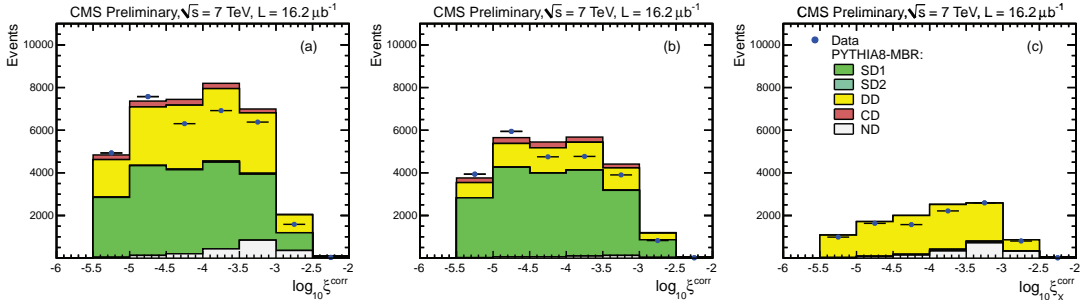}  
\end{center}
\vspace{-2em}
\caption{Detector-level distributions of $\xi$ for (a) the events after the $\eta_{min}>-1$ selection, and its subsamples corresponding to (b) the absence and (c) the presence of an energy deposit in {\sc castor}. The data are compared to predictions of the {\sc pythia8-mbr} simulation.}
\label{fig3}
\end{figure*}

\begin{figure}[!htb]
\vspace*{-1em}
\begin{center}
\includegraphics[width=0.66\textwidth]{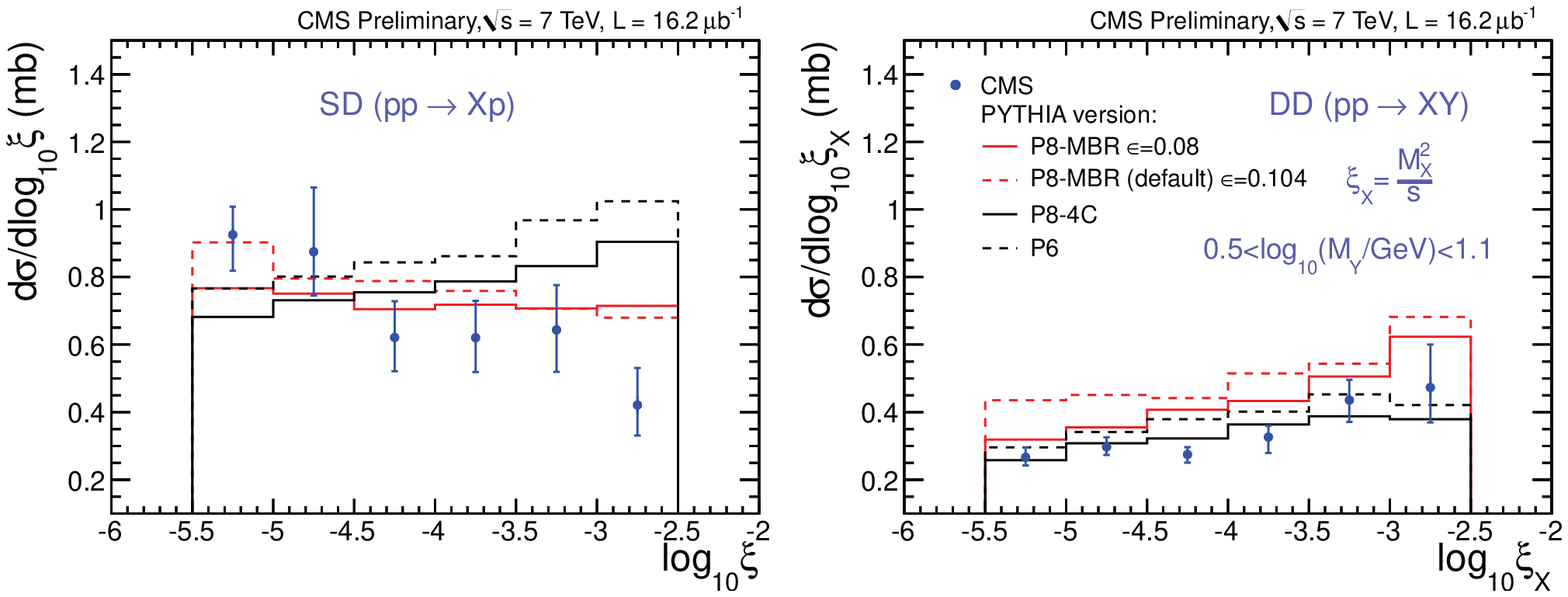}
\includegraphics[width=0.33\textwidth]{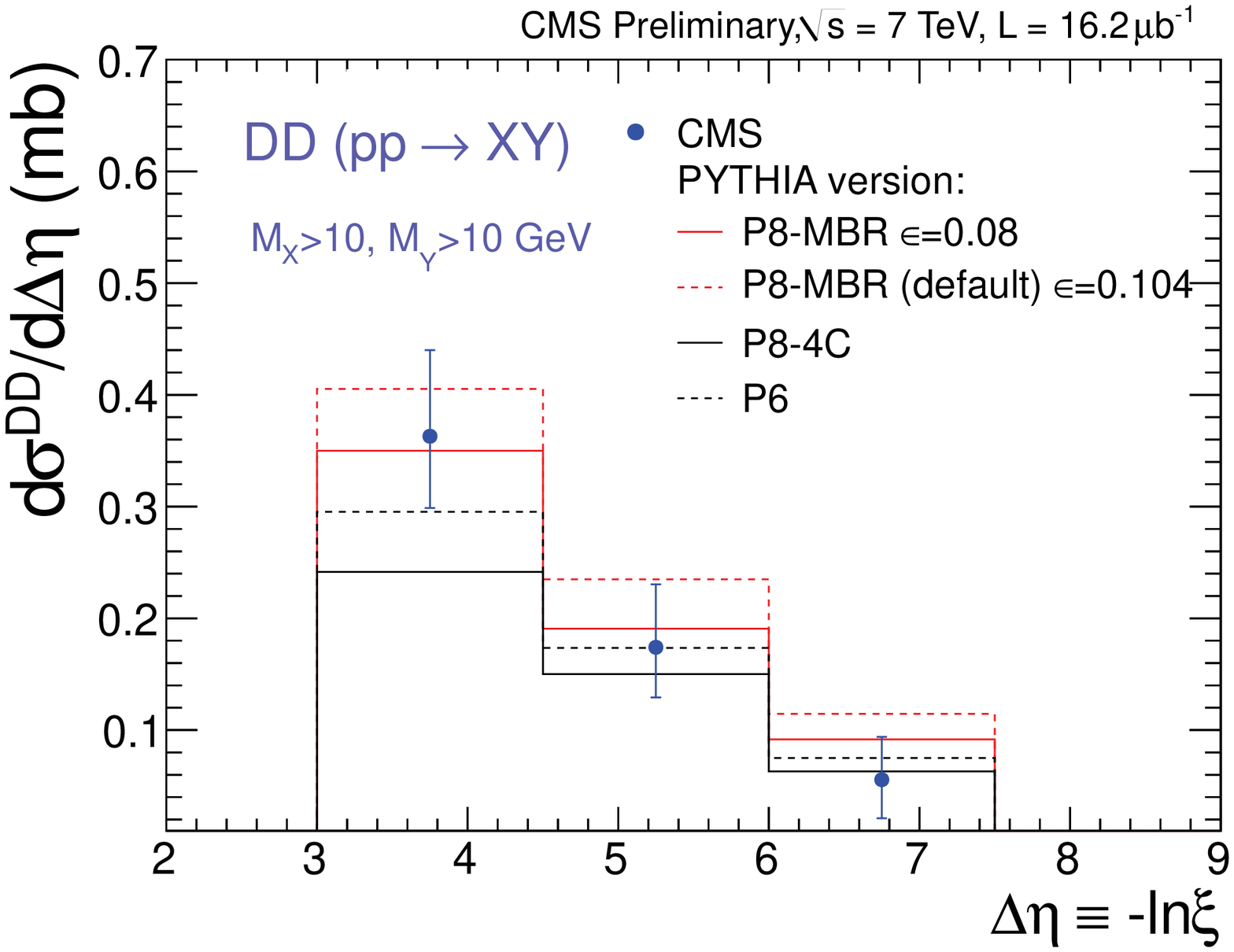}
\end{center}  
\vspace{-2em}
\caption{{\sc sd} (left) and {\sc dd} (middle) cross sections vs $\xi$, and {\sc dd} cross section vs $\Delta\eta$ (right), compared to {\sc pythia6}, {\sc pythia8-4c} and {\sc pythia8-mbr} MC. Errors are dominated by systematic uncertainties ({\sc hf} calorimeter energy scale, and hadronization and diffraction model).}
\label{fig4}
\end{figure}

The event sample after the $\eta_{min}>-1$ selection (Fig.~\ref{fig2}, middle) was used to extract SD and DD cross sections. The sample consists of approximately equal numbers of SD and DD events for which one of the dissociated masses is low and escapes detection in the central detector. Subsamples enhanced in SD and DD events were selected by requiring an absence or a presence of an energy deposit (above threshold) in the {\sc castor} calorimeter. The SD/DD separation with {\sc castor} is presented in Fig.~\ref{fig3}, showing the distribution of the variable $\xi$ calculated from all energy deposits in the detector. For SD events, $\xi$ approximates the incoming-proton momentum loss. These distributions were used to measure the differential SD cross section as a function of $\xi$, and the differential DD cross section as a function of $\xi_{X}={\rm M}^2_X/s$ for $0.5<log_{10}({\rm M}_Y/{\rm GeV})<1.1$ ({\sc castor} acceptance), after subtracting the background contribution to the signal (DD to SD and {\sc nd} to DD). 
Results are compared to MC models in Figs.~\ref{fig4} (left) and (middle), respectively. The predictions of {\sc pythia8-mbr} are shown for two values of the $\epsilon$ parameter of the Pomeron trajectory ($\alpha(t)=1+\epsilon+\alpha't$), $\epsilon = 0.08$ and $\epsilon = 0.104$. Both values describe the measured SD cross section within uncertainties, while the DD data favor the smaller value of $\epsilon$. The predictions of {\sc pythia8-4c} and {\sc pythia6} describe well the measured DD cross section, but fail to describe the falling behavior of the data~(see details in~\cite{softdiff}). The total SD cross cross section integrated over the region $-5.5<\log_{10}\xi<-2.5$ ($12 \lesssim M_X\lesssim 394$ GeV) was measured to be $\sigma^{SD}_{vis} = 4.27 \pm 0.04 (\rm{stat.}) ^{+0.65}_{-0.58} (\rm{syst.})$ mb (dissociation of either proton).

The event sample after the $\Delta \eta^{0}>3$ selection (Fig.~\ref{fig2}, right) was used to extract the differential {\sc dd} cross section as a function of the central-gap width, $\Delta\eta$. The cross section for $\Delta\eta>3$, M$_X>10$ GeV and M$_Y>10$ GeV is presented in Fig.~\ref{fig3} (right). The total {\sc dd} cross cross section integrated over this region was measured to be $\sigma^{DD}_{vis} = 0.93 \pm 0.01 (\rm{stat.}) ^{+0.26}_{-0.22} (\rm{syst.})$ mb.

\begin{wrapfigure}{r}{0.5\textwidth}
\vspace*{-2.5em}
\begin{center}
\includegraphics[width=0.4\textwidth]{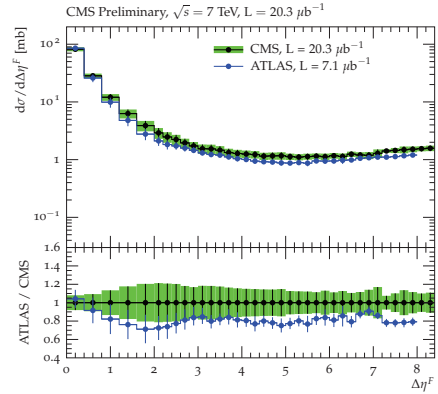}
\end{center}  
\vspace*{-2em}
\caption{$\mathrm{d}\sigma / \mathrm{d}\Delta\eta^{F}$ of events with a forward gap $\Delta\eta^F$ for stable particles with $p_{T} > 200$~MeV and $|\eta|<4.7$ compared with the {\sc atlas} measurement for $p_{T} > 200$~MeV and $|\eta|<4.9$~\cite{atlas}.}
\label{fig5}
\vspace*{-2em}
\end{wrapfigure}
\vspace*{-0.5em}

The inclusive differential cross section  $\mathrm{d}\sigma / \mathrm{d}\Delta\eta^{F}$ for events with a forward rapidity gap was measured, with $\Delta\eta^{F}$ defined as the larger of gaps within $\Delta\eta^{F} = max(4.7-\eta_{max},4.7+\eta_{min})$. A Bayesian-unfolded and fully corrected cross section for particles with $p_{T} > 200$ MeV and $|\eta|<4.7$ was extracted and compared to {\sc pythia6-z2*}~\cite{pythia6z2*}, {\sc pythia8-4c}, and {\sc pythia8-{\sc mbr}} ($\epsilon = 0.08$ and $0.104$) predictions. Figure~\ref{fig5} shows a comparison to the {\sc atlas} measurement~\cite{atlas}. The {\sc cms} measurement extends the {\sc atlas} result by 0.4 units of gap width. 
The two results are in agreement within the uncertainties, which at {\sc cms} are dominated by HF energy scale, hadronization, and diffraction model used.
%
\vspace*{0.5em}
\begin{footnotesize}

\end{footnotesize}
\end{document}